\documentclass[12pt]{article}
\usepackage{amsmath,latexsym}
\usepackage{graphicx}
\begin{document}

 \title{\Huge Particle Motion Around Tachyon Monopole }
 \author{ M.Kalam$^{\ddag}$, F.Rahaman$^*$ and  S.Mondal$^*$  }
\date{}
 \maketitle

 \begin{abstract}
Recently, Li and Liu  have studied global monoole of tachyon in a
four dimensional static space-time. We analyze the motion of
massless and massive particles around tachyon monopole.
Interestingly, for the bending of light rays due to tachyon
monopole instead of getting angle of deficit we find angle of
surplus. Also we find that the tachyon monopole exerts an
attractive gravitational force towards matter.
\end{abstract}

  \footnotetext{ Pacs Nos:  04.20 Gz, 04.50 + h, 04.20 Jb   
\\
 $*$Dept.of Mathematics, Jadavpur University, Kolkata-700 032,
 India:
                                  E-Mail:farook\_rahaman@yahoo.com\\
$\ddag$Dept. of Phys., Netaji Nagar College for Women, Regent
Estate, Kolkata-700092, India:

E-Mail:mehedikalam@yahoo.co.in

}
    \mbox{} \hspace{.2in}

\title{\Huge 1. Introduction: }

At the early stages of its evolution, the Universe has underwent
a number of  phase transitions. During the phase
 transitions, the symmetry has been broken. According to the
 Quantum field theory, these types of symmetry-breaking phase
 transitions produces topological defects [1]. These are namely
 domain walls, cosmic strings, monopoles and textures.  Monopoles are point like defects that may arise during phase transitions
 in the early universe. In particular , $ \pi_2 ( M ) \neq I $ ( M is the vacuum manifold ) i.e. M contains surfaces which can not be continuously
 shrunk to a point, then monopoles are formed [2]. \\
 A typical symmetry - breaking model is described by the
 Lagrangian,
\begin{equation}
               L =   \frac{1}{2}\partial_\mu \Phi^a \partial^\mu
               \Phi^a - V ( f )
         \label{Eq1}
          \end{equation}
Where $\Phi^a$ is a set of scalar fields, $ a = 1, 2, ….., N,
f=\sqrt{\Phi^a \Phi^a}$  and   $V( f )$ has a minimum at a non
zero value of  $f$. The model has $ 0(N)$ symmetry and admits
domain wall, string and monopole solutions for $ N = 1, 2 $ and $
3 $ respectively. It has been recently suggested by Cho and
Vilenkin(CV) [3,4] that topological defects can also be formed in
the models where $ V(f)$ is maximum at $f = 0$ and it decreases
monotonically to zero for
 $ f \rightarrow \infty $  without having any minima.
 For example,
 \begin{eqnarray*}
        V ( f ) = \lambda M^{4+n} ( M^n + f^n )^{-1}
 \end{eqnarray*}
 where $ M, \lambda $ and $ n $ are positive constants.
 This type of potential can arise in non-perturbative superstring models. Defects arising in these models
 are termed as " vacuumless defects ". Recently, several authors
 have studied vacuumless topological defects in alternative theory
 of gravity [5].

 Barriola and Vilenkin [6] were the pioneer who studied the gravitational effects of  global monopole. It was shown by considering only gravity that
the linearly divergent mass of global monopole has an effect
analogous to that of a deficit solid angle plus that of a tiny
mass at the origin [6]. Later it was studied by Harari and
Loust\`{o} [7], and Shi and Li [8] that this small gravitational
potential is actually repulsive. Recently, Sen [9] showed  in
string theories that classical decay of unstable D-brane produces
pressureless gas which has non-zero energy density. The basic
idea is that though the usual open string vacuum is unstable,
there exists a stable vacuum with zero energy density.This state
is associated with the condensation of electric flux tubes of
closed string [10]. By using an effective Born-Infeld action,
these flux tubes could be explained [11]. Sen also proposed the
tachyon rolling towards its minimum at infinity as a dark matter
candidate [10].  Sen have also analyzed the Dirac-Born-Infeld
Action on the Tachyon Kink and Vortex[12]. Gibbons actually
initiated the study of "tachyon cosmology". He took the coupling
into gravitational field by adding an Einstein-Hilbert term to the
effective action of the tachyon on a brane [13]. In the
cosmological background, several scientists have studied  the
process of rolling of the tachyon [14, 15].

Different kinds  of cold stars such as Q-stars have been proposed
to be a candidate for the cold dark matter [16-25]. A new class of
cold stars named as D-stars(defect stars) have been proposed by
Li et.al.[26]. Compared to Q-stars, the D-stars have a peculiar
phenomena, that is, in the absence of the matter field the theory
has monopole solutions,  which makes the D-stars behave very
differently from the Q-stars. Moreover, if the universe does not
inflate and the tachyon field T rolls down from the maximum of
its potential, the quantum fluctuations produced various
topological defects during spontaneous symmetry breaking. That is
why it is so crucial  to investigate the property and the gravity
of the topological defects of tachyon, such as Vortex [27], Kink
[28] and monopole, in the static space time. Recently, Li and Liu
[29] have studied gravitational field of global monopole of
tachyon.

In this paper, we will discuss the behavior of the motion of
massless and massive particles around Tachyon Monopole. We will
calculate the amount of deficit angle for the bending of light
rays. Also we will investigate the nature of gravitational field
of tachyon monopole towards  matters by using Hamilton-Jacobi method.\\

\pagebreak

\title{\Huge2. Tachyon Monopole Revisited: }

Let us consider, a general static, spherically-symmetric metric as
\begin {equation}
        ds^2=  A(r) dt^2 - B(r)dr^2 - r^2 (d\theta^2 + sin^2\theta d\phi^2 )
  \end{equation}

The Lagrangian density of rolling tachyon can be written in
Born-Infeld form as
\begin{eqnarray*}
        L = L_R + L_T
          = \sqrt{-g}\left[\frac{R}{2\kappa}-V(|T|)\sqrt{1-g^{\mu
          \nu}\partial_\mu T^a \partial_\nu T^a }\right]
 \end{eqnarray*}
where $ T^a $ is a triplet of tachyon fields, $ a = 1, 2, 3 $ and
$ g_{\mu\nu}$ is the metric coefficients. One can consider the
monopole as associated with a triplet of scalar field as
\begin{eqnarray*}
        T^a  =  f ( r )\frac{x^a}{r}
 \end{eqnarray*}
where $ x^a x^a = r^2 $. Now using the Lagrangian density, L, the
metric and the scalar field, the Einstein equations take the
following forms as
\begin{eqnarray*}
        \frac{1}{r^2}-\frac{1}{B}\left(\frac{1}{r^2}+\frac{B^\prime}{r
        B}\right) = \kappa T^0_0
 \end{eqnarray*}
\begin{eqnarray*}
        \frac{1}{r^2}-\frac{1}{B}\left(\frac{1}{r^2}+\frac{A^\prime}{r
        A}\right) = \kappa T^1_1
 \end{eqnarray*}
where the prime denotes the derivative with respect to r and
energy momentum tensor $ T^\mu_\nu $ are given by
\begin{eqnarray*}
         T^0_0 = V(f)\sqrt{1+\frac{f^{\prime 2}}{B}+\frac{2 f^2}{r^2}}
 \end{eqnarray*}
\begin{eqnarray*}
       T^1_1 = \frac{ V(f)(1+\frac{2 f^2}{r^2})}{\sqrt{1+\frac{f^{\prime 2}}{B}+\frac{2 f^2}{r^2}}}
 \end{eqnarray*}
\begin{eqnarray*}
       T^2_2 = T^3_3 = \frac{ V(f)(1+\frac{f^{\prime 2}}{B}+\frac{f^2}{r^2})}{\sqrt{1+\frac{f^{\prime 2}}{B}+\frac{2 f^2}{r^2}}}
 \end{eqnarray*}
and the rest are zero. So, the system depends on the tachyon
potential $ V(T)$. According  to Sen [9], the potential should
have an unstable maximum at $T = 0$ and decay exponentially to
zero when $T \rightarrow \infty $.

\pagebreak

One can choose the tachyon potential which satifies the above two
conditions as follows:
\begin{eqnarray*}
        V(f) =  M^4 \left(1+3\lambda f^4 \right)^{\frac{1}{6}} exp
        ( - \lambda f^4 )
 \end{eqnarray*}
where $M$ and $\lambda$ are positive constants.

In flat space-time, the Euler-Lagrange equation will take the
following form:
\begin{eqnarray*}
        \frac{1}{V}\left(\frac{dV}{df}\right) + \frac{2f}{r^2} =
        f^{\prime\prime}+\frac{2f^\prime}{r}-f^\prime \left[\frac{f^\prime f^{\prime\prime}+\frac{2f}{r^2}\left(f^\prime - \frac{f}{r}\right)}{1+f^{\prime
        2}+\frac{2f^2}{r^2}}\right]
 \end{eqnarray*}
and the energy density of the system can be  written as
\begin{eqnarray*}
         T^0_0 = V(f)\sqrt{1+f^{\prime 2}+\frac{2 f^2}{r^2}}
 \end{eqnarray*}
For the above mentioned tachyon potential, $ V(f)$ the
Euler-Lagrange equation has a simple exact solution
\begin{eqnarray*}
         f (r) =  \lambda ^ {- \frac{1}{4} } \left(\frac {\delta}{r}\right )
 \end{eqnarray*}
where $\delta = \lambda ^ {- \frac{1}{4} }$ is the size of the
monopole core and corresponding energy density becomes
\begin{eqnarray*}
         T^0_0 =  M^4 \left [ 1 + 3 \left ( \frac{\delta}{r}\right)^4 \right ]^{\frac{2}{3}} exp \left [-\left (\frac{\delta}{r}\right)^4 \right ]
 \end{eqnarray*}
Considering the Newtonian approximation, the Newtonian potential
can be written as
\begin{eqnarray*}
         \nabla^2 \Phi = \frac{\kappa}{2}(T^0_0 - T^i_i)
 \end{eqnarray*}
At $ r \gg \delta $,
\begin{eqnarray*}
          T^0_0 - T^i_i \simeq -2M^4 .
 \end{eqnarray*}
Therefore, the solution of the above equation is
\begin{eqnarray*}
           \Phi (r) \simeq - \frac{4 \pi M^4}{3 \lambda M_p^2 f^2 }
 \end{eqnarray*}
where $M_p$ is the Planck mass and the parameter M should
satisfies the condition $ M \leq 10^{-3} $ eV in order to avoid
conflicting present cosmological observations. The linearized
approximation applies for $ |\Phi(r)| \ll 1 $, which is
equivalent to $ f \gg \sqrt{\frac{4\pi}{3\lambda}}
\frac{M^2}{M_p} $.

Now, one can express the metric coefficients A(r) and B(r) as
\begin{eqnarray*}
            A(r) = 1 + \alpha(r),            B(r) = 1 + \beta(r).
 \end{eqnarray*}
Linearizing in $\alpha(r)$ and $\beta(r)$, and using the flat
space expression for $f(r)$, the Einstein equations becomes
\begin{eqnarray*}
         \frac{\alpha^\prime}{r} + \frac{\beta^\prime}{r} = \kappa M^4 \left(\frac{\delta}{r}\right)^4\left[1+3\left(\frac{\delta}{r}\right)^4\right]
         ^{-\frac{1}{3}} exp \left [-\left (\frac{\delta}{r}\right)^4 \right ]
 \end{eqnarray*}
and
\begin{eqnarray*}
         \alpha^{\prime\prime}+\frac{2\alpha^\prime}{r}  = -\kappa M^4 \left[2+3\left(\frac{\delta}{r}\right)^4\right]\left[1+3\left(\frac{\delta}{r}
         \right)^4\right]^{-\frac{1}{3}} exp \left [-\left (\frac{\delta}{r}\right)^4 \right ]
 \end{eqnarray*}
 After solving one can write the solution of the external metric as
\begin{equation}
       A(r) = \left(1-\frac{\kappa M^4}{3}r^2\right);  B(r) = \left(1+ \frac{\kappa M^4}{3}r^2 -\frac{\kappa M^4}{2\lambda
       r^2 }\right)
 \end{equation}

\title{\Huge3. The Geodesics: }

Let us now write down the equation for the geodesics in the
metric (2) . From
\begin{equation}
               \frac{d^2 x^\mu}{d\tau^2} + \Gamma^\mu_{\nu\lambda}
               \frac{d x^\nu}{d\tau}\frac{d x^\lambda}{d\tau}=0
               \end{equation}
we have
\begin{equation}
               B(r)\left(\frac{d r}{d\tau}\right)^2 = \frac{E^2}{A(r)} - \frac{J^2}{r^2} -  L
               \end{equation}
\begin{equation}
               r^2\left(\frac{d \phi}{d\tau}\right) =  J
               \end{equation}
\begin{equation}
                \frac{d t}{d\tau} = \frac{E}{A(r)}
               \end{equation}

where the motion is considered in the $ \theta  = \frac{\pi}{2}$
plane and constants E and J are identified as the energy per unit
mass and angular momentum, respectively , about an axis
perpendicular to the invariant plane $ \theta  = \frac{\pi}{2}$.
Here $\tau$ is the affine parameter and L is the Lagrangian having
values 0 and 1, respectively, for massless and massive
particles. \\
The  equation for radial geodesic ( $ J =0$):

\begin{equation}
        \dot{r}^2 \equiv \left(\frac{dr}{d\tau}\right)^2 = \frac{E^2}{A(r)B(r)} - \frac{L}{B(r)}
\end{equation}
Using equation(7) we get
\begin{equation}
         \left(\frac{dr}{dt}\right)^2 = \frac{A(r)}{B(r)} - \frac{A^2(r)L}{E^2B(r)}
\end{equation}

From equation(3), we can write
\begin{equation}
        \left(\frac{dr}{dt}\right)^2 = \left(1- \frac{\kappa M^4}{3} r^2\right)\left(1+ \frac{\kappa M^4}{3} r^2 - \frac{\kappa M^4}{2\lambda
               r^2}\right)^{-1} - \frac{L}{E^2}\left(1+ \frac{\kappa M^4}{3} r^2 - \frac{\kappa M^4}{2\lambda
               r^2}\right)^{-1}\left(1- \frac{\kappa M^4}{3} r^2\right)^2
\end{equation}
Expanding the expression binomially and neglecting the higher
order of $\kappa M^4$ ( as $\kappa M^4$ is very small ) we get
\begin{equation}
         \left(\frac{dr}{dt}\right)^2 = \left (1- \frac{2\kappa M^4}{3} r^2 + \frac{\kappa M^4}{2\lambda
               r^2}\right) - \frac{L}{E^2}\left(1- \kappa M^4 r^2 + \frac{\kappa M^4}{2\lambda
               r^2}\right)
\end{equation}
\title{\bf 3.1. Motion of Massless Particle ( L=0 ): }
In this case,
\begin{equation}
         \left(\frac{dr}{dt}\right)^2 =  \left(1- \frac{2\kappa M^4}{3} r^2 + \frac{\kappa M^4}{2\lambda r^2}\right)
\end{equation}
After integrating, we get
\begin{equation}
          \pm t = \int \frac{r dr}{\sqrt{\left(r^2- \frac{2\kappa M^4}{3} r^4 + \frac{\kappa M^4}{2\lambda
          }\right)}}
\end{equation}
This gives the $t - r $ relationship as
\begin{equation}
          \pm t =  - \frac{1}{\sqrt{ \frac{8\kappa M^4}{3}}}  sin^{-1} \left( \frac{1-\frac{4\kappa M^4}{3} r^2}{ \sqrt{1+\frac{4\kappa^2M^8}{3\lambda }}}\right)
\end{equation}
The $t - r$ relationship is depicted in Fig. 1.\\
\begin{figure}[htbp]
    \centering
        \includegraphics[scale=.8]{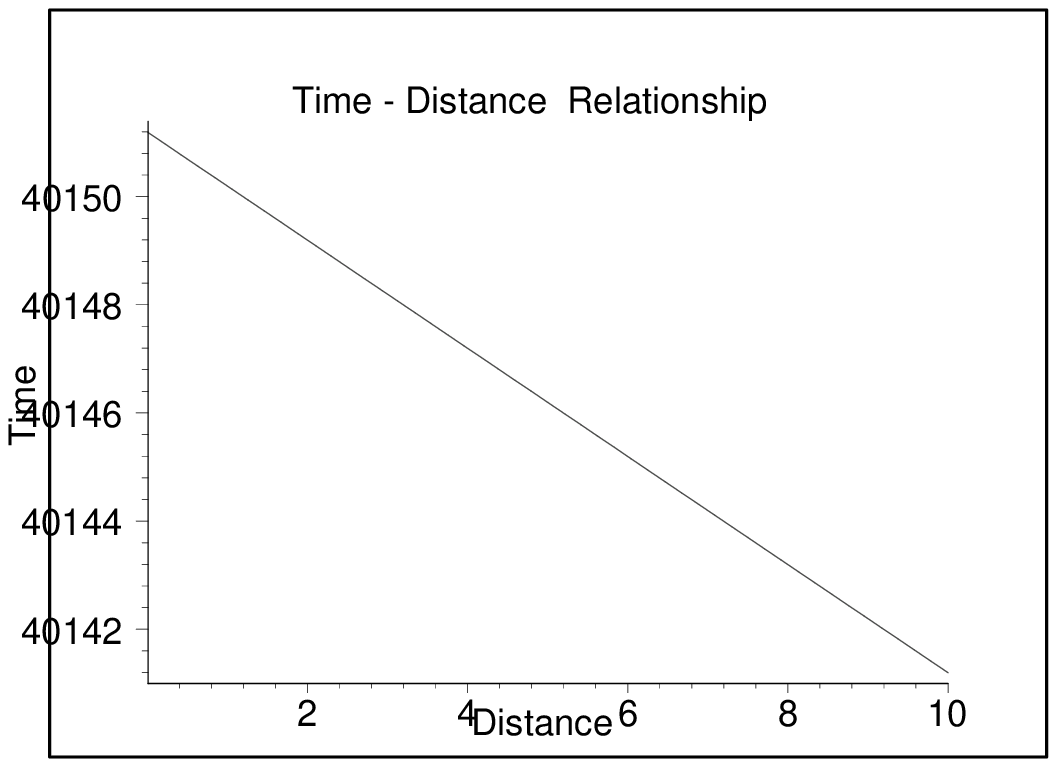}
    \caption{$t - r$ relationship for massless particle( choosing $\kappa M^4=573.95\times10^{-12}$, $\lambda =1$ ) }
    \label{fig:monopole}
\end{figure}

Again, from equation (8) we get
\begin{equation}
        \dot{r}^2 \equiv \left(\frac{dr}{d\tau}\right)^2 =  \frac{E^2}{A(r)B(r)}
\end{equation}

After integrating, we get
\begin{equation}
          \pm E \tau = \int  \sqrt{\left(1- \frac{\kappa M^4}{3} r^2\right)\left(1+ \frac{\kappa M^4}{3} r^2 - \frac{\kappa M^4}{2\lambda
               r^2}\right)}dr
\end{equation}
This gives the $\tau - r $ relationship as
\begin{equation}
          \pm E\tau =   \left( r + \frac{\kappa M^4}{4\lambda r}\right)
\end{equation}
( neglecting the higher order of $ \kappa M^4 $ ).

We show graphically (see Fig. 2 ) the variation of proper-time ($\tau$) with respect to radial co-ordinates (r) .\\

\begin{figure}[htbp]
    \centering
        \includegraphics[scale=.8]{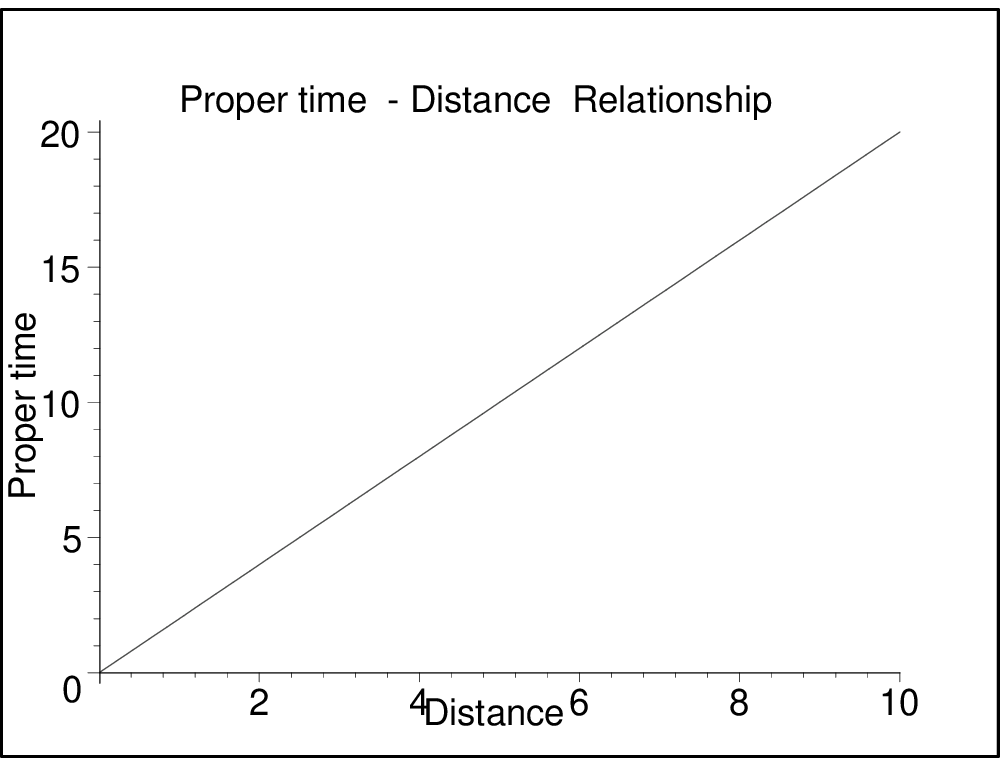}
    \caption{$\tau - r$ relationship for massless particle (  choosing  $ \kappa M^4 = 573.95\times10^{-12} $,           $ \lambda = 1, E = 0.5 $ )}
    \label{fig:monopole}
\end{figure}
\title{\bf 3.2. Motion of Massive Particles ( L=1 ): }
In this case,
\begin{equation}
  \left(\frac{dr}{dt}\right)^2 = \left(1- \frac{2\kappa M^4}{3}r^2+\frac{\kappa M^4}{2\lambda r^2}\right)-\frac{1}{E^2}\left(1- \kappa M^4 r^2+\frac{\kappa M^4}{2\lambda r^2}\right)
\end{equation}
After integrating, we get
\begin{equation}
          \pm t = \int \frac{E r dr}{\sqrt{\left(\kappa M^4-\frac{2\kappa M^4E^2}{3}\right) r^4 +\left(E^2-1\right)r^2 +
          \frac{\kappa M^4}{2\lambda}\left(E^2-1\right)   }}
\end{equation}
This gives the $t - r $ relationship as (see graphical Fig. (3))

 $
          \pm t =\frac{E/2}{\sqrt{\kappa M^4\left(1-\frac{2}{3}E^2\right)}} \ln  [
          2\sqrt{\left(\kappa M^4\left(1-\frac{2}{3}E^2\right)\right)\left(\kappa M^4\left(1-\frac{2}{3}E^2\right)r^4\right)+\left(E^2-1\right)r^2
          + \frac{\kappa M^4}{2\lambda}\left(E^2-1\right) }
\linebreak
          + 2\kappa M^4\left(1-\frac{2}{3}E^2\right)r^2 + \left(E^2-1\right)  ]
          $

\begin{figure}[htbp]
    \centering
        \includegraphics[scale=.8]{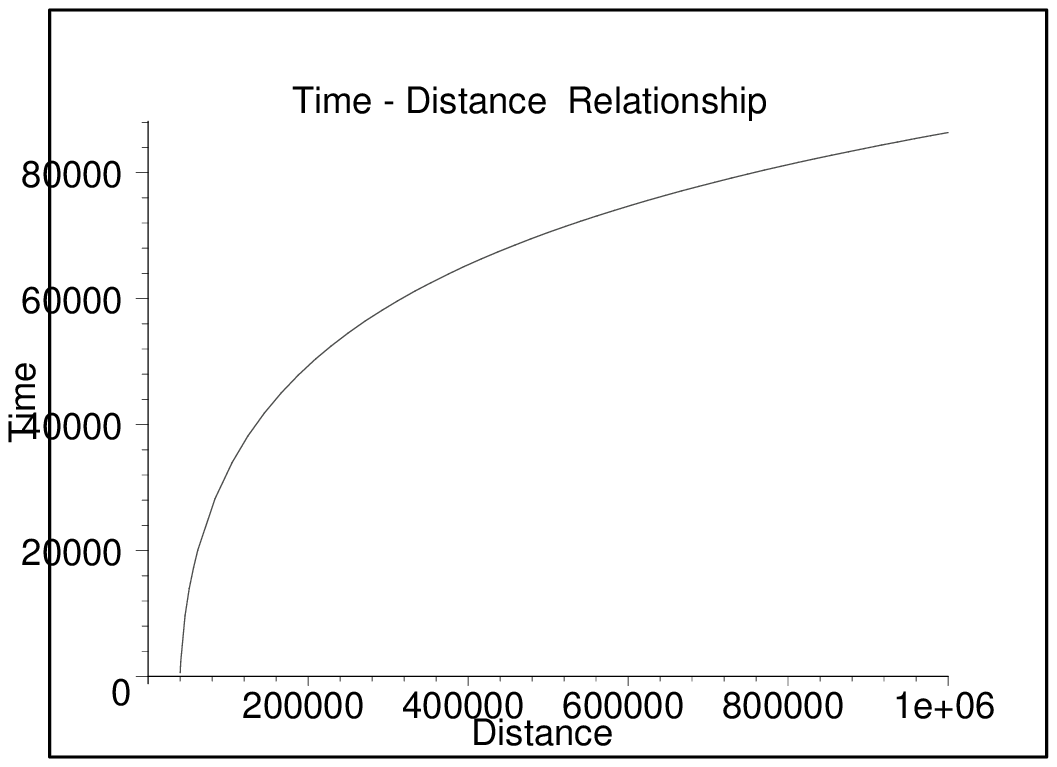}
    \caption{$t - r$ relationship  for massive particle( choosing  $ \kappa M^4 = 573.95\times10^{-12} $,           $ \lambda = 1, E = 0.5 $ )}
    \label{fig:monopole}
\end{figure}

Again, from equation (8) we get
\begin{eqnarray*}
        \dot{r}^2 \equiv \left(\frac{dr}{d\tau}\right)^2 =
        \frac{E^2}{A(r)B(r)} - \frac{1}{B(r)}
\end{eqnarray*}

Neglecting the higher order of $ \kappa M^4 $  , we get
\begin{eqnarray*}
          \pm \int  d\tau = \int \frac{\left(1-\frac{\kappa M^4}{4\lambda r^2}\right) dr }{\sqrt{E^2-1+\frac{\kappa M^4}{3}r^2}}
\end{eqnarray*}

This gives the $\tau - r $ relationship as
\begin{eqnarray*}
          \pm  \tau =  \sqrt{\frac{3}{\kappa M^4}} \cosh^{-1}\left[\frac{r}{\sqrt{\frac{3(1-E^2)}{\kappa M^4}}}\right]
          - \frac{(\kappa M^4)^{3/2}}{4\sqrt{3}\lambda (1-E^2)} \frac{\sqrt{r^2- \frac{3(1-E^2)}{\kappa M^4}}}{r}
\end{eqnarray*}

We show graphically (see Fig. 4 ) the variation of proper-time ($\tau$) with respect to radial co-ordinates (r) .\\

\pagebreak

\begin{figure}[htbp]
    \centering
        \includegraphics[scale=.8]{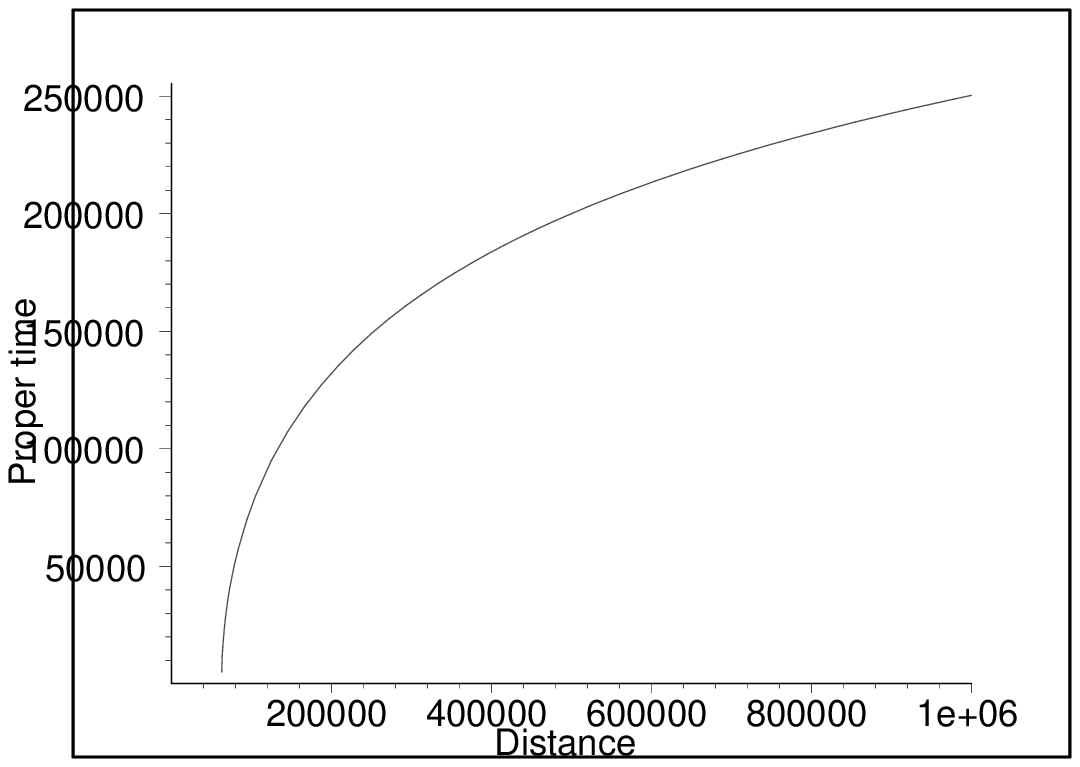}
    \caption{$\tau - r$ relationship for massive particle (  choosing  $ \kappa M^4 = 573.95\times10^{-12} $,           $ \lambda = 1, E = 0.5 $ )}
    \label{fig:monopole}
\end{figure}

\title{\Huge4. Bending of Light rays: }

For photons ( L=0 ), the trajectory equations (5) and (6) yield
\begin{equation}
        \left(\frac{dU}{d\phi}\right)^2 =  \frac{a^2}{A(r)B(r)} - \frac{U^2}{B(r)}
\end{equation}
where $ U = \frac{1}{r}$ and $ a^2 =\frac{E^2}{J^2}$.

Equation (20) and (3) gives
\begin{equation}
          \phi = \int \frac{\pm dU}{\sqrt{\left(a^2 + \frac{\kappa M^4}{3}\right)-\left(1- \frac{a^2 \kappa M^4}{2 \lambda}\right)U^2 }}
\end{equation}
( neglecting the higher order of $ \kappa M^4 $ and  the product of $ \kappa M^4 $   $\times$$U^4$ terms ). \\
This gives
\begin{equation}
          \phi =  \frac{1}{\sqrt{\left(1- \frac{a^2 \kappa M^4}{2 \lambda}\right)}} cos^{-1}\frac{U}{A}
\end{equation}
where $ A = \frac{a^2 +
\frac{\kappa M^4}{3}}{1-\frac{a^2\kappa M^4}{2\lambda}}$. \\
\begin{figure}[htbp]
    \centering
        \includegraphics[scale=.8]{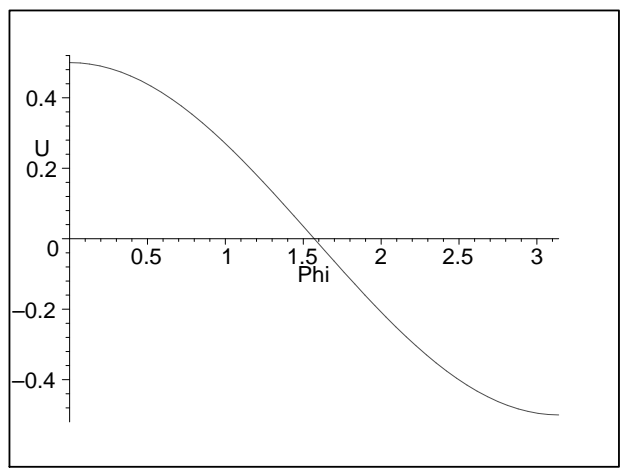}
    \caption{We Plot U vs. $\phi$ ( choosing  $\kappa M^4 = 573.95\times10^{-12}$, $\lambda = 1, a^2 = 0.5$) }
    \label{fig:monopole}
\end{figure}

For $ U \rightarrow 0 $, one gets
\begin{equation}
         2\phi =  \pi \left(1+ \frac{a^2 \kappa M^4}{4 \lambda}\right)
\end{equation}
 and bending comes out as
\begin{equation}
         \Delta\phi =  \pi - 2\phi = \pi- \pi\left(1+ \frac{a^2 \kappa M^4}{4\lambda}\right) = - \frac{a^2 \kappa M^4}{4\lambda} \pi
\end{equation}
which is nothing but angle of surplus[30].

\begin{figure}[htbp]
    \centering
        \includegraphics[scale=.8]{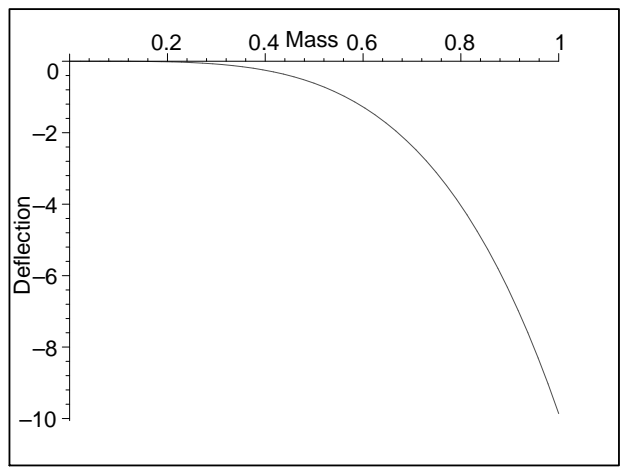}
    \caption{We plot Deflection vs. Mass ( choosing  $ \kappa  = 25.12 $, $ \lambda = 1, a^2 = 0.5 $ )   }
    \label{fig:monopole}
\end{figure}
\begin{figure}[htbp]
    \centering
        \includegraphics[scale=.8]{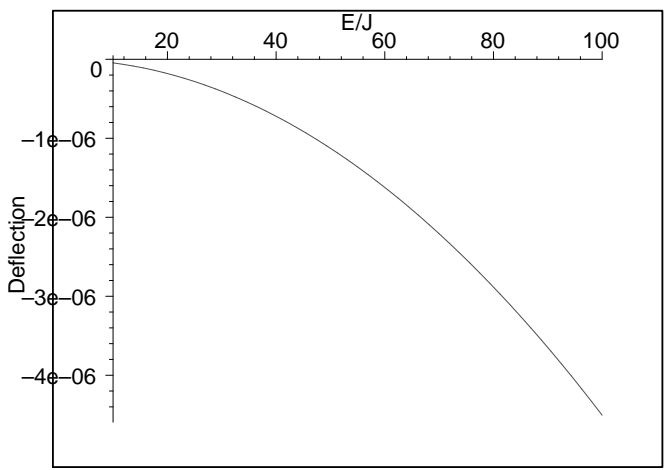}
    \caption{We plot Deflection vs. E/J ( choosing $\kappa M^4=573.95\times10^{-12}$,$\lambda=1$ )}
    \label{fig:monopole}
\end{figure}

\pagebreak

\title{\Huge5. Motion of test particle: }

Let us consider a test particle having mass $m_0$ moving in the
gravitational field of the tachyon monopole described by the
metric ansatz(2). So the Hamilton-Jacobi [ H-J ] equation for the
test particle is [31]
\begin{equation}
   g^{ik}\frac{\partial S}{\partial x^i} \frac{\partial S}{\partial x^k}+ m_0^2 = 0
 \end{equation}

   where $ g_{ik}$ are the classical background  field (2) and S is the standard Hamilton's
   characteristic function .

For the metric (2) the explicit form of H-J equation (25) is  [31]
\begin{equation}
   \frac{1}{A(r)}\left(\frac{\partial S}{\partial t}\right)^2 - \frac{1}{B(r)}\left(\frac{\partial S}{\partial
  r}\right)^2- \frac{1}{r^2} \left(\frac{\partial S}{\partial \theta}\right)^2-\frac{1}{r^2\sin^2\theta}\left(\frac{\partial S}{\partial \varphi}\right)^2
+  m_0^2 = 0
          \end{equation}

where  $A(r)$ and $B(r)$ are given in equation (3) .

In order to solve this partial differential equation, let us
choose the $H-J$ function $ S $ as [32]
 \begin{equation}
       S = - E.t + S_1(r) + S_2(\theta)  + J.\varphi
          \end{equation}
 where $E$ is identified as the energy of the particle and $J$
 is the momentum of the particle.

 The radial velocity of the particle is ( for detailed
calculations, see $ref.[32]$ )
\begin{equation}
         \frac{dr}{dt} = \frac{A(r)}{E\sqrt{B(r)}} \sqrt{\frac{E^2}{A(r)} +m_0^2 - \frac{p^2}{r^2} }
          \end{equation}
where $p$ is the separation constant.

The turning points of the trajectory are given by
$\left(\frac{dr}{dt}\right) = 0 $ and as a consequence the
potential curve are
\begin{equation}
         \frac{E}{m_0} = \sqrt{A(r) \left(\frac{p^2}{m_0^2r^2} - 1\right)} \equiv V(r)
          \end{equation}
In a stationary system, $ E $ i.e. $ V(r)$ must have an extremal
value. Hence the value of $r$ for which energy attains it
extremal value is given by the equation
\begin{equation}
         \frac{dV}{dr} =   0
          \end{equation}

\pagebreak

 Hence we get
\begin{equation}
         \frac{2\kappa M^4}{3} r^4 = \frac {2 p^2}{m^2}
         \Rightarrow r = \left(\frac{3 p^2}{\kappa M^4 m^2}\right)^{\frac{1}{4}}
          \end{equation}
So this equation has at least one positive real root. Therefore,
it is possible to have bound orbit for the test particle i.e. the
test particle can be trapped by the tachyon monopole. In other
words, the tachyon monopole exerts an attractive gravitational
force towards matter.

\title{\Huge6. Concluding remarks: }

In this paper, we have investigated the behavior of a massless and
massive particles in the gravitational field of a tachyon
monopole. The tachyon monopole, in compare to the ordinary
monopole, are very diffuse objects whose energy distributed at
large distances from the monopole core, their space-time is vastly
different from the ordinary monopole. The figures (1) and (2)
indicate that the nature of ordinary time and proper time for the
massless particle in the gravitational field of tachyonic monopole
is opposite to each other. Here, one can see that ordinary time
decreases with increase of radial distance where as the proper
time increases with increase of radial distance. Figures (3) and
(4) show that in case of massive particle, the ordinary time and
proper time have the same nature.
 According to Li and Liu
[29], tachyon monopole has a small gravitational potential of
repulsive nature, corresponding to a negative mass at origin. In
the analysis of the bending of light rays, we get angle of
surplus instead of angle of deficit. So, we may conclude that it
has a property of short range repulsive force. From eqn.(31), we
see that $ r = \frac{1}{M}\left(\frac{3 p^2}{\kappa
m^2}\right)^{\frac{1}{4}} $ i.e.  $r$   would be very large as
$M$ is very small,  in other words, particle can be trapped at a
large distance from the monopole core.  This implies tachyon
monopole would have effect on particles far away from its core.
That means tachyon monopole has a long range gravitational field
which is sharply contrast to ordinary monopole.

 {  \Huge Acknowledgments }

          F.R. is thankful to DST , Government of India for providing
          financial support.  MK has been partially supported by
          UGC, Government of India under MRP scheme. \\


\end{document}